\documentclass[journal]{IEEEtran}

\ifCLASSINFOpdf
\else
\fi

\usepackage{setspace, amsmath, amssymb, url, lscape, subfigure, algorithmic, multirow, pslatex, listings, verbatim, alltt, amsfonts, wrapfig, boxedminipage, color, cite} %,bookmark}
\usepackage[vlined,linesnumbered,ruled,boxed]{algorithm2e}
\usepackage[dvips]{graphicx}
\usepackage{epsfig}
\usepackage{mathtools}
\usepackage{xcolor}
\newcommand{\qed}{\nobreak \ifvmode \relax \else
      \ifdim\lastskip<1.5em \hskip-\lastskip
     \hskip1.5em plus0em minus0.5em \fi \nobreak
      \vrule height0.75em width0.5em depth0.25em\fi}

\newcommand{\eg}{{\it e.g., }}
\newcommand{\etal}{{\it et~al., }}
\newcommand{\ie}{{\it i.e., }}

\newcommand{\comments}[1]{}
\newcommand\hl{\bgroup\markoverwith
  {\textcolor{yellow}{\rule[-.5ex]{2pt}{2.5ex}}}\ULon}
\newcommand{\RNum}[1]{\uppercase\expandafter{\romannumeral #1\relax}}

\begin{document}

\title{Edge Computing for User-Centric Secure Search on Cloud-Based Encrypted Big Data}

\author{\IEEEauthorblockN{Sahan Ahmad\IEEEauthorrefmark{1},
SM Zobaed\IEEEauthorrefmark{1},
Raju Gottumukkala\IEEEauthorrefmark{2}, and
Mohsen Amini Salehi\IEEEauthorrefmark{1}}

\IEEEauthorblockA{\IEEEauthorrefmark{1}School of Computing \& Informatics \\
}
\IEEEauthorblockA{\IEEEauthorrefmark{2}Informatics Research Institute\\
University of Louisiana at Lafayette\\
Lafayette,Louisiana-70504, USA\\Email: \{sahan.ahmad1,sm.zobaed1, raju, amini\}@louisiana.edu
%Lafayette,Louisiana-70504, USA\\
}
}

% The paper headers
%\markboth{Manuscript submitted to IEEE Transactions on Security}
%{Shell \MakeLowercase{\textit{et al.}}: Bare Demo of IEEEtran.cls for Journals}

\maketitle

\begin{abstract}
Cloud service providers offer a low-cost and convenient solution to host %big large
unstructured data. 
However, cloud services act as third-party solutions and do not provide control of the data to users. This has raised security and privacy concerns for many organizations (users) with sensitive data to utilize cloud-based solutions. User-side encryption can potentially address these concerns by establishing user-centric cloud services and granting data control to the user. Nonetheless, user-side encryption limits the ability to process (\eg search) encrypted data on the cloud. Accordingly, in this research, we provide a framework that enables processing (in particular, searching) of encrypted multi-organizational (\ie multi-source) big data without revealing the data to cloud provider. Our framework leverages locality feature of edge computing to offer a user-centric search ability in a real-time manner. In particular, the edge system intelligently predicts the user's search pattern and prunes the multi-source big data search space to reduce the search time. The pruning system is based on efficient sampling from 
the clustered big dataset on the cloud. For each cluster, the pruning system dynamically samples appropriate number of terms based on the user's search tendency, so that the cluster is optimally represented. We developed a prototype of a user-centric search system and evaluated it against multiple datasets. Experimental results demonstrate 27\% improvement in the pruning quality and search accuracy.

%We create samples for each small portion of encrypted cluster data based on user search.   Based on user search interest, we create an unencrypted sample for encrypted cluster data in the edge which can represent the topic of the cluster. The representation of encrypted cluster data is critical as we can prune the clusters based on its topic while performing search operation and maintain real-time semantic search on encrypted big data. We also introduced the optimal number of the element that needs to represent a cluster. Moreover, we also introduce how to choose the best element that can represent a small portion of a cluster which will eventually help to represent the overall topic of the cluster. This representation improves the search precision and also improves the performance. 
\end{abstract}


\section{Introduction}\label{sec:intro}

Cloud services provide flexible and scalable solutions to host and process big data. The number of businesses and individuals using cloud solutions for big data is skyrocketing and the volume of data stored on cloud is expected to surpass 44 ZB in near future~\cite{infoot}.  
 %Alleviates the maintenance cost for own datacenter, availability of network device, technical advancement to use cloud service for an application makes cloud service an reasonable solution to host and process data. 
However, data privacy concerns still preclude many businesses, specially those with sensitive and confidential data (\eg criminal reports or financial documents), to utilize cloud services. One noticeable recent data privacy incident was \emph{Panama paper leaking} in which around $11.5$ million documents of the Panama paper were leaked. The documents contained detail financial and attorney-client information for more than $214,488$ offshore bank entities including actors, politicians, athletes, and businessmen \cite{panama01}. Thus, data privacy has remained a major concern in using cloud services to host data~\cite{javanmard2015tsc}.

Organizations are in need of solutions that securely store their large document sets in the cloud and enable operations (\eg real-time search) on them without revealing the documents to unauthorized  and malicious users. With the prevalence of thin-clients (\eg smart-phones), users expect lightweight solutions that impose a minimal load on the user device. Particularly for big data, sophisticated search methods, such as semantic search~\cite{guha2003semantic}, are desired. The reason is that in big data a concept can be potentially expressed in various ways. For a given search query, \emph{semantic search} can find all documents that contain phrase(s) conceptually or ontologically related to the search query~\cite{woodworth2018s3bd}. %One approach to achieve security is to encrypt client data on cloud servers. However, this approach does not protect documents against internal attackers~\cite{zobaedbig}. 

The motivation of this study comes from multiple law-enforcement departments (\eg detectives, criminologists, and sheriffs) desire to perform semantic search on a large document set of criminal reports~\cite{lenss}, stored in a cloud environment. These documents are confidential and access to them must be limited to the law-enforcement officers who need to search the data using their hand-held devices in a real-time manner. As such, the law-enforcement governing body desires a solution to protect confidentiality of the big document set against external or internal attackers in the could, while maintaining the semantic search ability.

User-side encryption~\cite{woodworth2016s3c,salehi2014reseed} is an approach to achieve data privacy against internal and external attacks~\cite{kaaniche2014secure,litwin2011privacy}. In this approach, documents are encrypted with the user key, before outsourcing them to the cloud and only users possessing the encryption key can decrypt the documents. However, the problem in user-side encryption is losing the ability to perform operations (\eg semantic search) on the encrypted documents. Another problem, which is particularly prominent for big datasets is to maintain the real-time response in processing the encrypted documents.

%the problem (how to optimally prune clusters.)
Searchable encryption systems (\eg \cite{song2000practical}) enable search over encrypted documents. These systems generally maintain an index to map the keywords with their associated document~\cite{pham2018survey}. However, with big datasets, the index size can become extensively large~\cite{chou2011parallel} and the real-timeness of the system is affected. To make searchable encryption systems scalable, solutions are provided to partition the encrypted keywords of the index structure into several clusters based on the topical relatedness of keywords~\cite{woodworth2016s3c}~\cite{clust2019}. Then, for a given search query, a \emph{pruning} method is used to limit the search space only to clusters relevant to the query~\cite{woodworth2018s3bd}. However, the current searchable encryption systems for big data introduce two challenges:
\begin{enumerate}
\item  Although pruning method reduces the search time, it impacts the search precision because of covering only a subset of the indexed keywords.
In fact, pruning is achieved based on \emph{abstracts} that include sample keywords from each one of the encrypted clusters. The abstracts are utilized on the user side in an unencrypted form to navigate the search only to relevant clusters. To improve search precision, the sampled keywords of an abstract should precisely capture the topics of a corresponding cluster. In addition, sampling quantity for an abstract (\ie number of keywords in an abstract) must reflect user tendency in searching within a particular cluster.

\item The search systems introduce a significant overhead to the user's thin-client. Several resource-intensive components of the search system (\eg creating and updating abstracts, document encryption, keyword extraction, and pruning) reside on the user side that overwhelmes the user's thin-client. 

%To reduce the underlying overheads from user-side, we propose to embrace edge computing to reduce client side computation. 
\end{enumerate}

To address these challenges, in this study, we propose a method to dynamically sample from each encrypted cluster and form a summary (\ie abstract) structure that qualitatively represent topics of each cluster and quantitatively reflect the user search interest to the cluster. As the keywords in the clusters are encrypted and we cannot infer anything about their semantics, we base the sampling method on the keywords and phrases already searched by the users. Our proposed method functions based on Markov Chain model~\cite{dias2007latent} to analyze users' search tendency and nominate keywords whose inclusion in the abstracts can improve the search quality. Quantity of keywords in each abstract is determined based on the variety of search queries navigated to a given cluster. 

Edge/fog computing has emerged to fill the gap between client machines and remote cloud datacenters. As we move from the cloud to user premises,  communication latency decreases but trustworthiness increases~\cite{patel2017using}. There are edge computing solutions, such as Fortivault produced by Fortinet ltd~\cite{FortinetLTD}, that act as a gateway between the user premises and cloud and its goal is to secure accessing to cloud datacenters. In such solutions, because edge machines are on the user's premises, both the client and edge machines are deemed trustworthy.
We leverage the edge computing machines to eliminate the overheads of creating abstracts and other search components (\eg keyword extraction) from the user side device (thin-client). In addition, edge computing helps in reducing redundant processing needed to generate abstracts for users within the same department and with similar search interests. % However, the data needs to be protected from malicious access while it is on the cloud.

In summary, contributions of this study are as follows:
\begin{itemize}
 \item Developing an edge computing framework that enables lightweight semantic search on encrypted unstructured big data, hosted in cloud, in a real-time manner. 
 
 \item Proposing a theory to efficiently sample from encrypted clusters. The theory enables sampling each cluster to an abstract whose quality of its elements represents topics of the cluster and quantity of it elements reflects the user search tendency.
 
 %\item Proposed a theory for sampling cluster data intelligently based on user search history, semantic distance, user interest and data volume in the cluster to create sample that can represnt all the topics of data held by the cluster.
 %\item Proposed system architecture which can support multi-organization search schema.
 \item Evaluating and analyzing the search quality of the proposed framework against existing and established methods.
\end{itemize}

The rest of the paper is organized as follows: In Section~\ref{sec:relatedWork}, we review recent studies undertaken in this area. Then, Section~\ref{sec:architecture} provides an overview of the system architecture and explains functionality of each component. Section~\ref{sec:method} describes the proposed theory of our solution. Section ~\ref{sec:security} provides security analysis for edge tier. Section~\ref{sec:exp} evaluates the performance of the proposed theory based on real-world datasets. Finally, Section~\ref{sec:conclusion} concludes the paper and explains avenues of future work.

\section{Related Work}\label{sec:relatedWork}

In this section, we review research studies recently undertaken in cloud-based search solutions and position our contribution with respect to them. More specifically, we consider four related areas, namely user behavior model in information retrieval, user search pattern analysis, sampling from encrypted data, Secure Semantic Search. 

\noindent\textbf{User Behavior Model in Information Retrieval.}
To guess user intention while searching a query in the system, there has been extensive research work to analyze user behavior. Different models have been applied to analyze user behavior as an example reinforcement learning ~\cite{athukorala2016beyond}, Neural Model ~\cite{mitra2017neural}. The final goal is to retrieve highly related documents on top ranks based on user search.  The user behavior can vary in a different domain. As an example, Palotti \etal~\cite{palotti2016users} analysis logs from different medical sources and shows that the search behavior varies from medical professionals to laypeople.  Though our research objective does not limit to a specific domain, it contains the encrypted unstructured document sets on the cloud server and the data set  are divided on different cluster based on relevancy. The precision of the retrieved document highly dependent on finding out the relevant cluster after a search is performed by the user.  As, we are able to extract only data volume information from the cloud server, we consider analyze user behavior is the appropriate option to project the data characteristics of a cluster.

\noindent\textbf{User Search Pattern Analysis.}
In the information storage and retrieval area, several studies have been undertaken on analyzing the users' search behavior (\eg \cite{joachims2017accurately}). Dias and Vermunt~\cite{dias2007latent} proposed a model-based clustering approach based on user search pattern to understand what type of information online market users demand in their interaction with the websites. They used Markov Chain to categorize a series of web pages visited by users and predict the probability that a user would visit a web page at a given time \textit{t}. Ai \etal~\cite{ai2017characterizing} analysis email search behaviors and observed that user tend to find out specific keyword in the mail search rather than a generalized term. Benevenuto \etal~\cite{benevenuto2009characterizing} analyzed user surfing behaviors in online social networks to understand the frequency and the time duration people spend on social media. They analyzed the streamed data to identify patterns of use in social networks. Rose \etal~\cite{rose2004understanding} presented a framework to understand the goals of a user searching on the web. This research works focus on predicting users' behavior on the web. Alternatively, we are inspired to incorporate the idea of user behavior analysis on encrypted unstructured documents and clusters built on them. Another difference is in our objective, which is to utilize user search behavior knowledge to improve the accuracy of our search system.

\noindent\textbf{Sampling from Encrypted Data.}
Sample size determination for plain-text data is well established~\cite{sim2018can,chow2017sample}. Many research works focus on domain-specific sampling (\eg for genomes data) and determining the minimum sample size based on the purpose of analysis in that domain~\cite{QualitativeResearch}. Eng~\cite{eng2003sample} proposed five parameters that can determine the appropriate sample size from a population. These parameters are namely, estimated measurement variability, desired statistical power, significance criterion, and the shape of planned statistical analysis. Minimum sample size determination depends on the way the data is collected and the nature of the data. To the best of our knowledge, there is no research work targeting sample size determination for encrypted clustered data, which is our focus in this research. 

\noindent\textbf{Secure Semantic Search}
There have been a few research works that expand the idea of semantic search over encrypted data~\cite{woodworth2016s3c,woodworth2018s3bd}. Research works based on a fuzzy keyword search on encrypted data have been undertaken to overcome the semantic search on encrypted data. Liu \etal \cite{liu2011fuzzy} proposed a secure search schema based on fuzzy matching of keywords in which the value of matching between two keywords varies between zero and one. They proposed a method to reduce the index size, hence, reducing the search time. Fu \etal \cite{fu2018semantic} also presented a method based on the semantic relationship between concepts in encrypted datasets. S3BD~\cite{woodworth2018s3bd} is a semantic search system based on user-side encryption and pruning technique on a cluster of encrypted data to reduce the search time.% S3BD introduces sampling from cluster keywords to navigate the search to appropriate clusters. The sampling from each cluster is based the centroid and a constant number of (ten) keywords of the cluster with the highest number of document associations. In S3BD, the whole sample construction, query processing, and pruning are taken care on the client-end, making the solution infeasible for thin-client devices.%
Our research uses Woodworth's user-end encryption method, however, we leverage edge computing paradigm that enables lightweight client-side and provides us more control over the searched data to improve the search precision. %While uploading the document, we extracted keywords from the document. In the upload process, we send encrypted term with frequency information to the cloud server.

\section{Edge-Based Architecture for Secure Search Over Big Data}\label{sec:architecture}

%\subsection{Overview of System Architecture}

\begin{figure}
\centering
\includegraphics[scale=0.35]{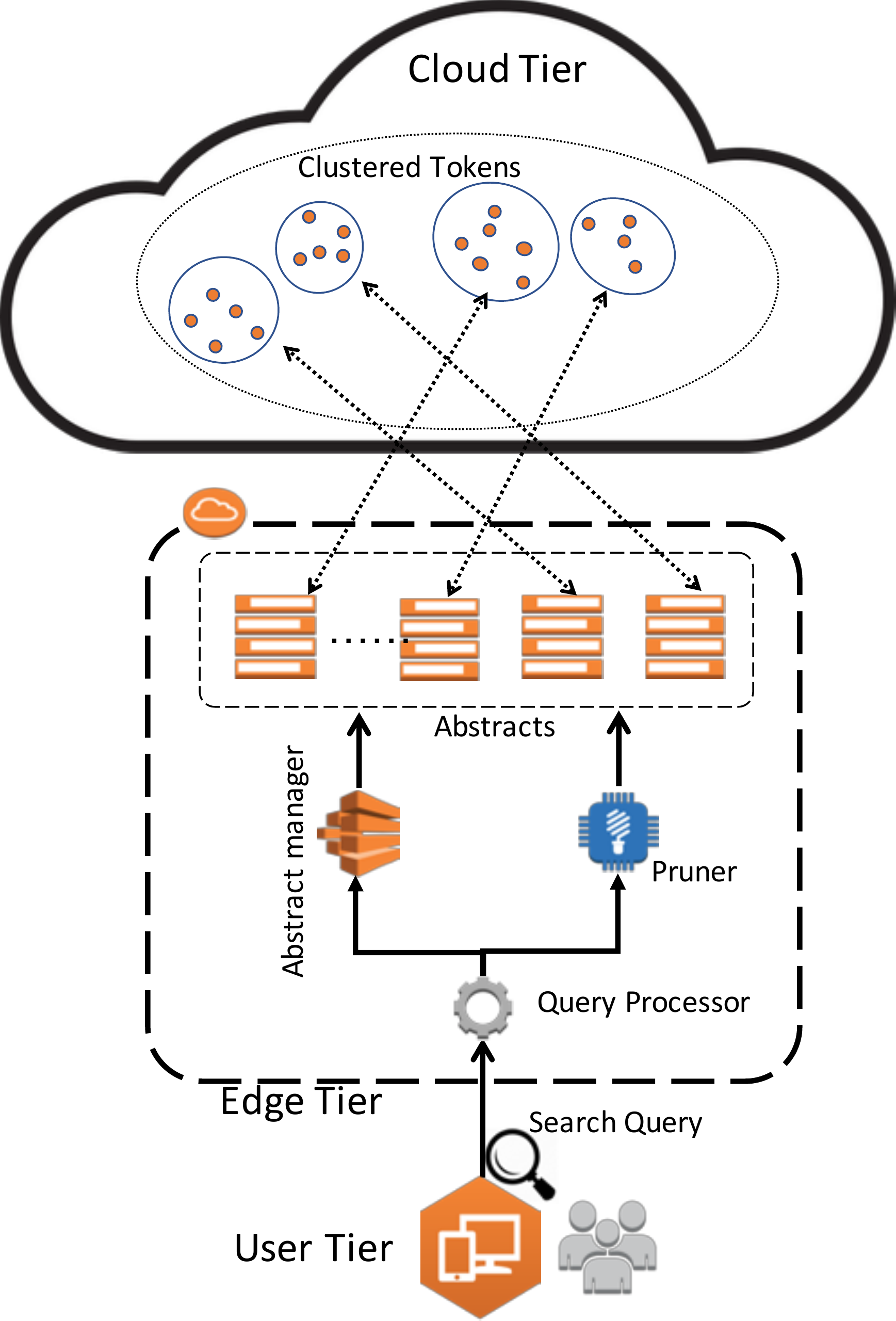}
\caption{Bird's eye-view of Edge-Based Architecture for Secure Search System Architecture and major processes.}
\label{fig:systemArchitecture}
\end{figure}

Our secure search system is composed of three tiers: \emph{user tier},  \emph{edge tier}, and \emph{cloud tier}. 
Figure~\ref{fig:systemArchitecture} provides a bird's-eye-view of primary components of each tier and major processes of the system. The details of each tier is as follows:

\textbf{User Tier.} User tier is composed of a lightweight interface to search over the encrypted data on the cloud. It is hosted on the users' thin-clients that assumed to be trusted. User's search query is sent to the edge tier. The search result is sent back to the user from the cloud provider in form of a list of documents ranked based on their relevance. %The user application is lightweight as the only job to the client to send the request and receive the response from the cloud storage.

\textbf{Edge Tier.}  The edge tier is a computing facility on the users' premises (organization), hence, considered to be trusted. %The security analysis of edge tier is presented on section ~\ref{sec:security}. 
\emph{Query Processor} module receives the search query, pre-processes it by removing stop-words, stemming, splitting multi-phrase queries to capture partially-matching documents~\cite{woodworth2016s3c}. Then, the query is augmented to encompass the conceptual and antological semantics of the query~\cite{woodworth2016s3c}. \emph{Abstracts} is a set of lists of unencrypted keywords that each one represents a sample of clustered keywords reside on the cloud. There is a one-to-one association between each list (\ie abstract) and a cluster. 
\emph{Abstract Manager} is a module that monitors the search queries to identify users' search interest within an organization and accordingly updates the abstracts. This module operates periodically and in an offline manner. For a given search query, \emph{Pruner} module determines the most related abstracts and limits the search operation only to the corresponding clusters. The edge tier (particularly, Abstract Manager and Pruner modules) is the main module explored in this paper (see Section~\ref{sec:method}). 

\textbf{Cloud Tier.} It is a third-party cloud provider with large storage and computing services. We consider the cloud provider as \emph{honest but curious}, hence, cannot be trusted. The cloud storage service is used to store encrypted documents and to cluster their encrypted keywords. Clustering is performed on the cloud based on the keywords extracted from documents and their co-occurrence frequency \cite{woodworth2018s3bd}. These information are extracted during the document upload process. 

To avoid searching the entire dataset and perform effective pruning, clusters need to be topic-based, so that semantically related terms are co-located in the same cluster~\cite{endert2012semantics}. Because of encryption, extracting the topic of terms is not feasible, hence, topic-based clustering is carried out based on statistical semantics~\cite{wang2017clustering}. Particularly, in S3BD ~\cite{woodworth2018s3bd}, for topic-based clustering of encrypted terms, terms' frequency of co-occurrences across documents is considered as the statistical semantics. In this case, we define \emph{semantic relatedness} between two given terms based on the intersection of their appearances across documents. 

Once the semantic relatedness between terms is defined, K-means algorithim~\cite{kmeans} can be used to cluster the terms. For that purpose, k encrypted terms must be chosen as the centroids. Term $\omega$ is chosen as a centroid if the number of documents uniquely associated with $\omega$ is more than the number of documents it shared with other terms. Once the centroids are chosen, other terms are clustered with the centroid that they share the maximum semantic relatedness. Further details about clustering can be found in ~\cite{woodworth2018s3bd}.

The cloud computing service is used to search over a subset of (\ie pruned) clusters, upon receiving a search request. Then, the results are ranked based on the relevance to the search query, before sending them to the user. Our objective, in this study, is to protect the data stored in the cloud from illicit access by internal and/or external attackers.

\section{Sampling from an encrypted cluster}\label{sec:method}
\subsection{Overview}
Recall that there is a one-to-one association between an abstract and a cluster. In fact, each abstract is a representation of its corresponding cluster that resides on the edge tier, can be quickly compared against the search query, and navigates the search only to relevant clusters (\ie pruning). 

Accurate pruning and choosing relevant clusters strongly depends on how representative the corresponding abstracts are. In fact, quantity and quality of sampled terms from a cluster determine representativeness of the abstract structure. The \emph{quantity} refers to the sample size from the corresponding cluster, whereas the \emph{quality} of the abstract depends on how the chosen terms conceptually express the cluster. 

As for the optimal abstract (sample) size, we should note that, too small sample size impacts representativeness of the abstract, inaccuracy of pruning, and subsequently, the search precision. In contrast, too large abstract size increases the overhead and impacts the real-time quality of the search operation. 

The search precision is impacted by the abstract quality (\emph{sampling quality}). For a given search query, based on the semantic relatedness between abstracts and the query, the Pruner determines which clusters on the cloud must be excluded or included in the search operation. Therefore, each term  chosen for an abstract should semantically represent a portion (\ie sub-topic) of the corresponding cluster and the union of such chosen terms in an abstract should represent topics of the entire cluster. Moreover, the abstracts should reflect the changes occur to the dataset (\eg because of document addition or removal). As such, the Abstract Manager should have the flexibility of dropping sampled terms that lose their importance over time and replacing them by more representative ones. 

According to Figure~\ref{fig:systemArchitecture}, the Abstract Manager module in the edge tier is in charge of constructing and populating abstracts. As the terms in the clusters are encrypted, the Abstract Manager cannot use them to build abstracts. Therefore, we propose to infer the abstract items from users' search queries and their metadata. In particular, we propose to determine the quality and quantity of the abstract terms based on the following factors: \textit{user search pattern}, \textit{users' interest to a cluster}, and \textit{number of cluster terms} (aka cluster size). In the rest of this section, we elaborate on the impact of these factors on the quality and/or quantity of terms that form an abstract. We also describe calculating these factors for a given cluster and determining the most representative terms for its corresponding abstract.
%We utilize the above mention factors information to add or replace the in the sample while maintaing the optimal quality of abst
%For adding or replacing an item in the sample, we consider a semantic replacable radius value. To calculate semantic radius value, we use the value returned by the above mention three factors. The qualification of any term to add or replace dynamically in the sample, we use the weight return by the Markov Chain Model for the term.  Through this way, we eventually maintain the optimal number of terms in the sample dynamically with respect to time. 

\subsection{Analyzing Users' Search Pattern}
%Sampling encrypted terms in clusters without decrypting them is infeasible, as their true meaning (semantic) is lost. To overcome this problem,
Abstract Manager considers users' search pattern as an accessible metadata and leverages that for quality sampling that can semantically project the topics of the cluster. %Specifically, by analyzing the metadata, we are able to infer the terms in the clusters and create a sample . 
In addition, analyzing search pattern enables us to learn about search queries that potentially appear in near future. This knowledge can be leveraged to improve the precision of pruning and subsequently the search operation. %Note that search pattern analysis occurs in the edge tier that is considered to be at the user's premises, hence, trustworthy. 

To predict users' search queries, we develop a method based on Markov Chain model, because it can describe a sequence of possible events where the probability of each event depends only on the state attained in the previous event. For that purpose, we create an adjacency graph that contains the sequence of search queries made by users to a particular cluster and leverage that to predict search queries that can potentially appear in the near future.

\begin{figure}
\centering
\includegraphics[width=0.45\textwidth]{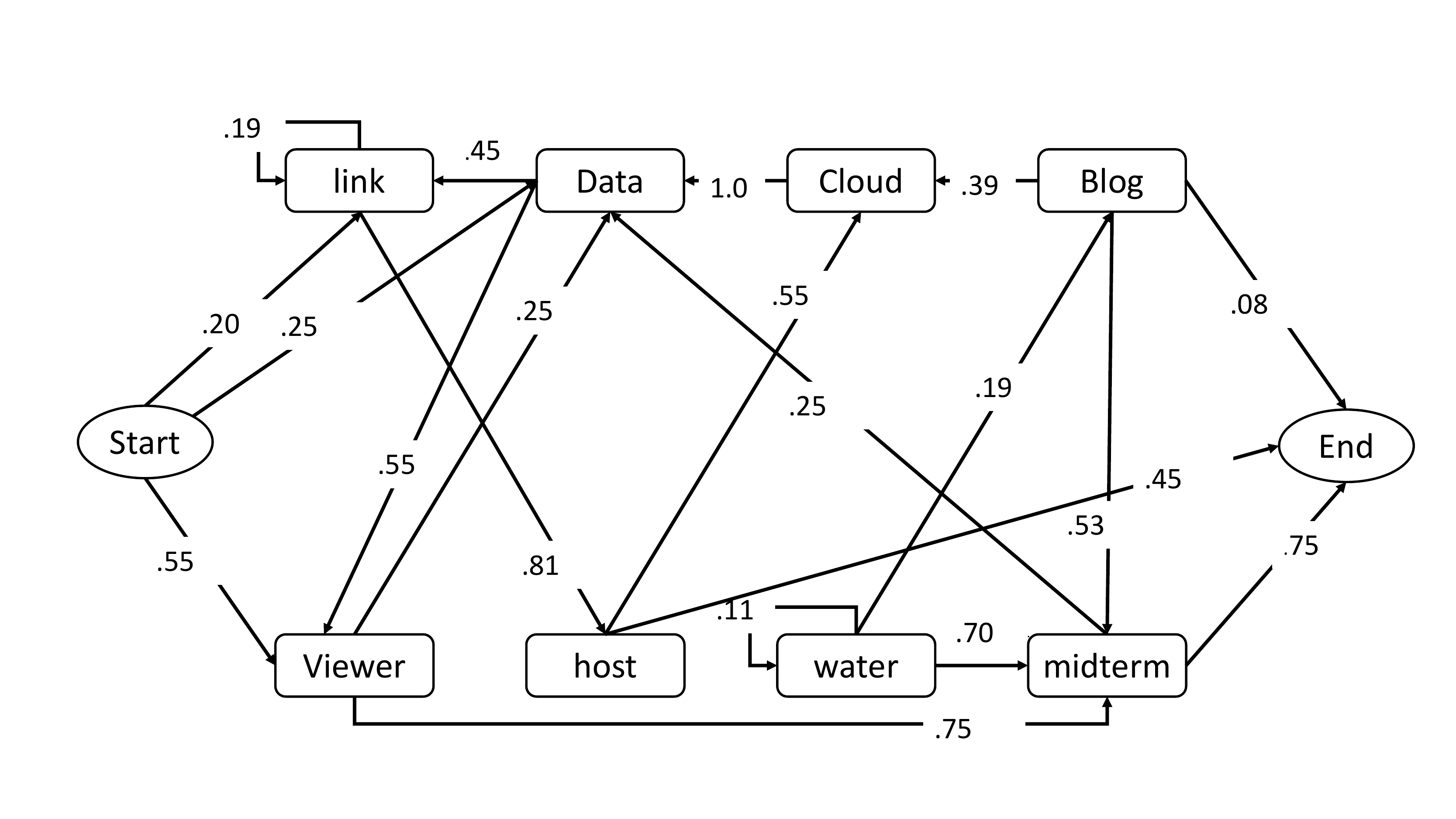}\vspace{-0.5cm}
\caption{An overview of user search pattern analysis using Markov Chain.
The states represent user search terms and the arrows and their weights represent the transition probability from one state to another. The ovals denote the start and end of an interval.}
\label{fig:userSearchPattern}
\end{figure}

Let $W$ be the set of $m$ searched terms (we have $W:\{w_1, w_2, ....,w_m\}$) appear in the system during a certain time interval and hit a particular cluster $c$. Elements of $W$ form the states of a Markov Chain. Figure~\ref{fig:userSearchPattern} represents Markov Chain in form of a graph where each vertex $w_k$ shows a search term and each edge $(w_k,w_j)$ shows the probability that term $w_j$ is searched right after $w_k$. This probability between adjacent vertices is termed \emph{transition probability} and denoted as $P_{kj}$. Then, the \emph{n-step} transition probability between $w_i$ and $w_j$, denoted $w_{ij}(n)$, is defined as the compound probability of searching term $w_j$, after $n$ steps from searching $w_i$. Equation~\ref{eq:markov} formally shows how $w_{ij}(n)$ is calculated. 
\begin{equation}\label{eq:markov}
w_{ij}(n)= \sum_{k=1}^{m}w_{ik}(n-1)\cdotp P_{kj}
\end{equation}

The \emph{state probability} of term $w_i$, denoted $P_{i}^s$, is defined as the likelihood of the term being searched in future. The initial value for the state probability of $w_i$ is calculated based on the ratio of searching frequency for $w_i$ to the total number of searched terms in cluster $c$. 

For the set of all terms $w_i \in W$, the state probabilities form a matrix with $m$ elements and transition probabilities form a $m\times m$ matrix. 
Then, the probability of searching terms in future is obtained by multiplying the state probability matrix with transition probability matrix iteratively, until the state probability matrix converges. The converged state probability matrix also indicates the appropriate terms that can be chosen for the abstract (called \emph{qualified terms}). Specifically, the state probability of term $w_i$ can be used to add or replace an term in the abstract structure of that cluster. %A term with a higher weight can replace an existing term with a lower weight. 

\subsection{Users' Interest to a Cluster}
To perform accurate pruning and navigate search to relevant clusters, in determining the quantity of terms in each abstract, we should consider the users' interest to  the corresponding clusters. The clusters that are targeted more frequently (\ie higher users' interest) should be sampled more granularly (\ie have larger abstracts) and vice versa.

For a given encrypted cluster $i$, the number of search queries hit the cluster show its popularity and can be leveraged to infer the topics covered by that cluster. In particular, cluster popularity can be vary because the cluster covers several topics and/or because users are interested to the few topics of that cluster. Popularity of a cluster is measured with respect to the number of queries other clusters receive. In practice, distribution of queries across clusters is unbalanced~\cite{woodworth2018s3bd}. Therefore, to measure popularity of a given cluster $i$, we use the ratio of deviation of number of queries hit cluster $i$ to the mean number of queries each cluster should ideally receive. Let $q_i$ the number of queries hit cluster $i$ and $\bar{q}$ the average number of queries each cluster should receive. Then, \emph{popularity} of cluster $i$ is defined as: $\sigma_i= (q_i-\bar{q})/\bar{q}$. 

Let $0\leq \delta_{i}(a,b)\leq 1$ represent the semantic similarity between queries $a$ and $b$ that hit encrypted cluster $i$. Then, the average semantic similarity of all queries that hit the cluster (denoted $\bar{\delta_i}$) show variations of topics contained by cluster $i$. In particular, dissimilar search queries (\ie $\bar{\delta_i} \rightarrow 0$) imply that the cluster terms cover several topics. Alternatively, $\bar{\delta_i}\rightarrow 1$ indicates that the cluster terms are concentrated on a few topics. 

We define users' interest to cluster $i$, denoted $\beta_i$, based on their interest to the topics of that cluster. As such, both the popularity of the cluster ($\sigma_i$) and the average semantic similarity of queries hit the cluster ($\bar{\delta_i}$) can be used to measure the users' interest to cluster $i $. Therefore, we calculate $\beta_i$ as: $\beta_i = 1/(\delta_i + \sigma_i$). 
%We note that, it is possible that encrypted clusters are unbalanced. That is, some clusters include many terms while others contain just few terms. As such, the cluster popularity ($\sigma_i$) can get biased to the more populated clusters.
% To mitigate such impact on the users' interest to cluster $i$, we use the idea of harmonic average of $\sigma_i$ and $\bar{\delta_i}$ to reduce the impact of large outliers. Therefore, $\beta_i$ is calculated based on Equation~\ref{eq:UI}.

% \begin{equation}\label{eq:ASD}
% \delta = \frac{ \sum_{k=1}^{n}\sum_{j=1}^{n}SemanticSimilarity(term_{kj})}{n}
% \end{equation}

\subsection{Cluster Size}
For a given cluster $i$, the number of terms in that cluster (\ie cluster size) is positively correlated with the quantity of terms need to be sampled from that cluster. The relationship between population and sample size is not linear and it varies based on different application and data nature \cite{QualitativeResearch}. Hence, to determine the number of terms for the abstract representing cluster $i$, in addition to users' interest to cluster $i$, we consider the number of terms in the cluster (denoted as $\gamma_i$). 

%Let $\bar{\gamma}$ represent the expected cluster size in a well-balanced clustering and let $c_i$ represent the actual size of cluster $i$. We define \emph{balance} of cluster $i$, denoted $\gamma_i$, as the ratio of deviation from the expected cluster size in a well-balanced clustering (\ie $\gamma_i= (c_i-\bar{\gamma})/\bar{\gamma}$).

\subsection{Procedure to Add or Remove Terms in Abstract Structures}
Initially, abstracts are populated with equal number of (\eg ten) terms, chosen based on the number of document association in each cluster~\cite{woodworth2018s3bd}. Then, while users issue search queries, the edge tier collects metadata and refines the abstracts' terms based on the factors mentioned in the previous part and by using the following main steps: 
\begin{enumerate}
 \item Analyzing users' search pattern and determine if a searched term is qualified to be placed in the abstracts or not.
 \item If the term is qualified, then the appropriate abstract is selected.
 \item Within the selected abstract, it is determined either to add the qualified term or replace it with an existing term.
\end{enumerate}
 In the rest of this section, we elaborate on each one of these steps.

\noindent\textbf{1. Determining qualified terms to be placed in the abstracts.}
Successful search queries issued by users are examined against the Markov Chain model explained in Section \ref{eq:markov} to learn their potential use in future.
The terms that have a significant user interest can be considered as an abstract term and are called \emph{qualified terms}.

\noindent\textbf{2. Selecting appropriate abstract.}
To select an appropriate abstract for a qualified term, semantic similarity of the qualified term and existing members of each abstract is measured based on Wu Palmer method~\cite{wupalmer}. The abstract that offers the highest semantic similarity with the qualified term is selected as the target abstract. Ties are broken by selecting the abstract that is more often used, when the qualified term is searched.

\noindent\textbf{3. Determining adding or replacing the qualified term in the selected abstract.}
Assume abstract $i$ is selected for a qualified term $w$. Term $w$ is added to abstract $i$, if $w$ introduces a topic that is not covered by existing terms in the abstract. Otherwise, $w$ replaces the most semantically similar term exist in the abstract $i$, only if it has a higher probability to appear in future than the existing abstract term. Accordingly, the decision to add or replace $w$ in abstract $i$ is made based on the cluster size($\gamma_i$) and the user interest to the corresponding cluster ($\beta_i$). That is, the higher the cluster size and the more user interest for a cluster imply a larger abstract size. 

We define \emph{semantic radius} for abstract $i$, denoted $SR_i$, as the threshold semantic similarity to add or replace terms in abstract $i$. %Low semantic radius implies high semantic similarity and vice versa. 
For a qualified term $w$, if it is within the semantic radius of an existing term $t$ of abstract $i$ (\ie $\delta_i(w,t)\geq SR_i$), then it implies that the topic $w$ represents is already being covered by $t$. In this circumstance, the competition between $w$ and $t$ is resolved based on the future appearance probability of the two terms that is calculated based on the Markov Chain model. Alternatively, if $w$ is not in the semantic radius of any term in the selected abstract $i$ (\ie $\delta_i(w,\forall t\in i)<SR_i$), it implies that the topic has not been covered by the existing abstract terms. Hence, it has to be added to abstract $i$. The value of $SR_i$ is calculated based on Equation~\ref{eq:SR}.
\begin{equation}\label{eq:SR}
SR_i = \frac{1}{\delta_i + \sigma_i + log(\gamma_i)}
\end{equation}

%We calculate $SR_i$ for $i$th cluster as $SR_i = \beta_i + \gamma_i $. 
\begin{figure}
\centering
\includegraphics[width=0.3\textwidth]{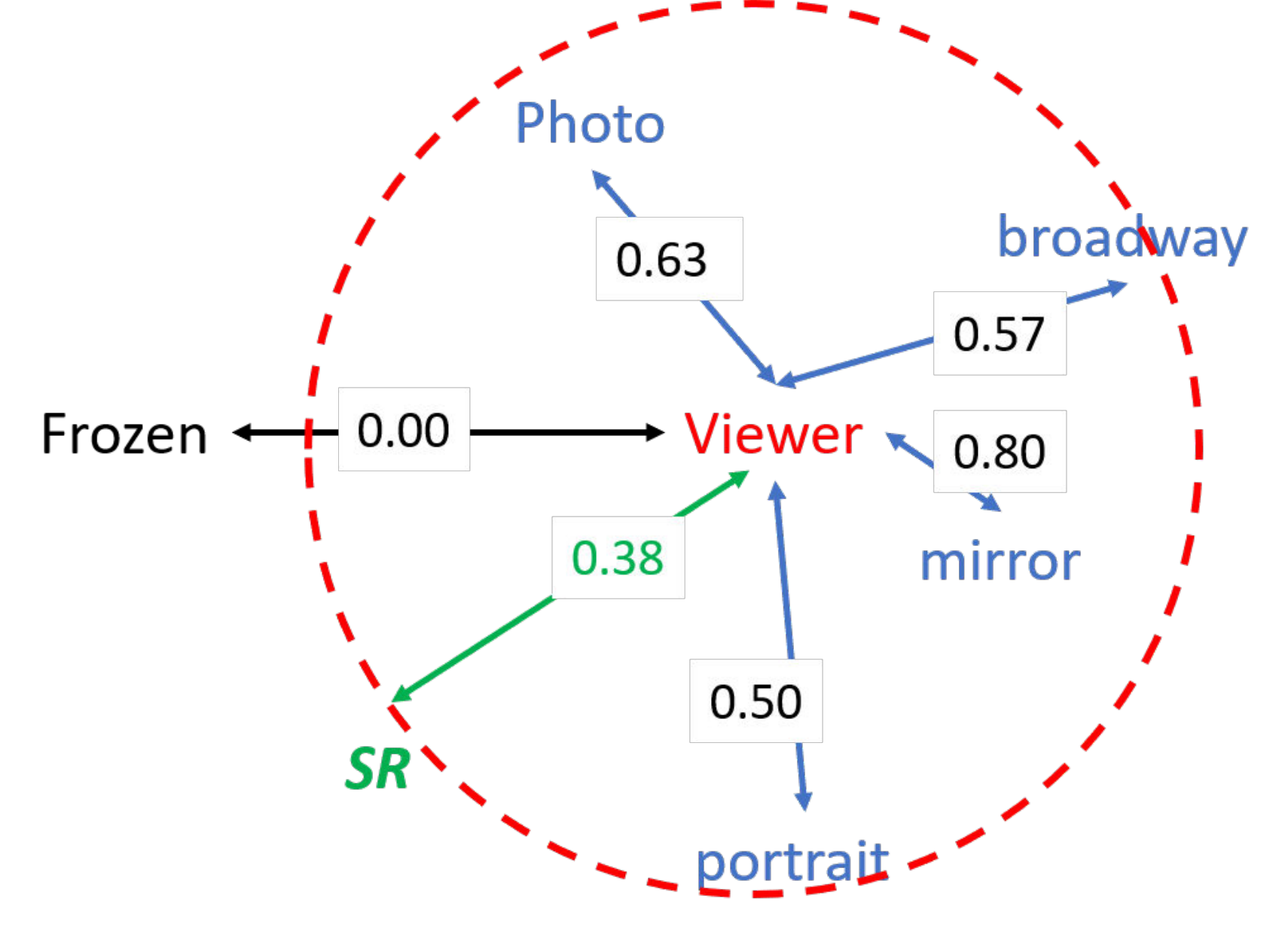}
\caption{Example of semantic radius (SR=0.38) for term \emph{Viewer} in an abstract. Blue terms are within and black term (\emph{Frozen}) is outside of the semantic radius. Numbers show semantic similarity between the \emph{Viewer} and other terms.}
\label{fig:SR2Image}
\end{figure}

Figure~\ref{fig:SR2Image} visually shows and example of a semantic radius, where the value of $SR$ is assumed to be 0.38. In this example, \textit{Viewer} is an existing abstract term and \textit{Frozen, Portrait, Broadway, Mirror,} and \emph{Photo} are considered as qualified terms. Semantic similarity of all the qualified terms with \textit{Viewer} are shown. The calculated semantic similarity of \textit{Frozen} and \textit{Viewer} is $0.00$. Hence, \textit{Frozen} represents a topic that is not covered by \emph{Viewer} and its being added to the abstract. However, semantic similarity of other terms with \textit{Viewer} fall within the semantic radius that implies topical overlapping. In this case, the term with the highest future appearance probability becomes a term in the abstract and other terms are discarded.
It is noteworthy that the terms lose users' interest over time become less competitive and are replaced with qualified terms. %In this manner, our method ensures that the terms in the abstract have a significant contribution.%We can think a bunch of little clusters that make the bigger cluster. If we can represent the small cluster, collectively we can represent the whole cluster.

\section{Security Analysis}\label{sec:security}
Using clouds, data owner loses his/her full control over the outsourced data. Our threat model considers internal and/or external attacks to privacy of the outsourced cloud data and the goal is to preserve the data privacy. Our solution preserves data privacy by maintaining documents, their metadata (clustered terms), communication, and processing only in encrypted format. Hence, the attacker can learn nothing about documents and cluster tokens, once they are out of the user (trusted) premises.

Edge computing paradigm is predominantly used for low-latency communication~\cite{patel2017using}. From data control perspective, however, this paradigm brings multi-layer data control architecture to cloud-based storage solutions. From security perspective, edge computing resides at the user's premises, hence, can be considered trusted. For instance, Fortinet ltd.~\cite{FortinetLTD} has developed a secure edge platform, called Fortivault, that is deployed on a trusted hardware at the user's premises to facilitate secure access to the cloud~\cite{salehi2014reseed}. Using edge-based platforms, the further data get from the users' machine, the less privacy it is provided with. However, as edge resources still reside in user's premises, they offer more control and privacy in compare to third-party clouds. In this study, our aim is to leverage the secure property of edge computing to enable accurate, real-time, and yet lightweight search solution while maintaining big data privacy.

We note that, even though edge is considered trusted, it does not have access to user's private key. It only contains the abstract structures to facilitates the search. In case, the edge is attacked, the attacker can only learn about the topics of each cluster and user's queries. However, documents and clusters terms remain private and secure.

% To execute an application-specific customized computation, researchers propose an intelligence at the edge of the network, which has more computational capabilities with respect to client device~\cite{patel2017using}. Recently, Amazon launches Greengrass which is an edge-analytic software solution where the developer can push small analytical capabilities. 

\section{Performance Evaluation}\label{sec:exp}

\subsection{Experimental Setup and DataSet}

For performance evaluation, we implemented a prototype of the proposed system in the context of S3BD. We used two datasets, namely Amazon Common Crawl Corpus (ACCC) \cite{comcrawl} and Request For Comments (RFC) \cite{rfcdataset}, that have distinct characteristics and volumes. ACCC is $\approx$ 150 terabytes, contains web contents, and is not domain-specific. RFC is domain-specific and contains topics about Internet and network protocols. It includes 6298 documents and the size is $\approx$  247 MB. %We choose these datasets so that we can analyze our approach competence on different data volume and data variation. 
All the experiments were carried out on 10-core 2.8 GHz E5-2680v2 Xeon processor with 64 GB memory. On this machine, we created three Virtual Machines (VMs) to represent client, edge, and cloud tiers. 

To simulate users' search behavior and evaluate our system, we synthesized 1,000 benchmark queries for the RFC and 2,000 for the ACCC dataset. To synthesize queries from these datasets, we used Maui~\cite{Maui}, a term extractor tool, to obtain 15 keywords from a subset of documents in each datasets. Then, we concatenate three keywords extracted from each document to generate one search query. We used 70\% of the benchmark for learning users' search pattern and the rest of 30$\%$ queries for evaluating our proposed system in identifying the right cluster.

%To evaluate the efficiency of the abstract structures, we run search queries in batch and measure the accuracy of finding relevant cluster for each search query  as the document retrieval process highly dependent on searching on the relevent cluster. We compare our work with S3BD to measure the quality of our abstract in case of navigating relevant cluster for the search queries. To visualize our sampling quality, we present cluster word formation along with the terms in the abstract. It shows how the terms in the cluster are distributed around the terms in the abstract.To evaluate choosing the three-factor and each of their contribution to the proposed sampling method, we perform growth testing. To evaluate our approach performance against big data, we measure the search completion time against ACCC dataset. We also examined the storage overhead incurred in the edge tier for analyzing user search data.
\begin{figure}
\centering
\includegraphics[width=0.30\textwidth]{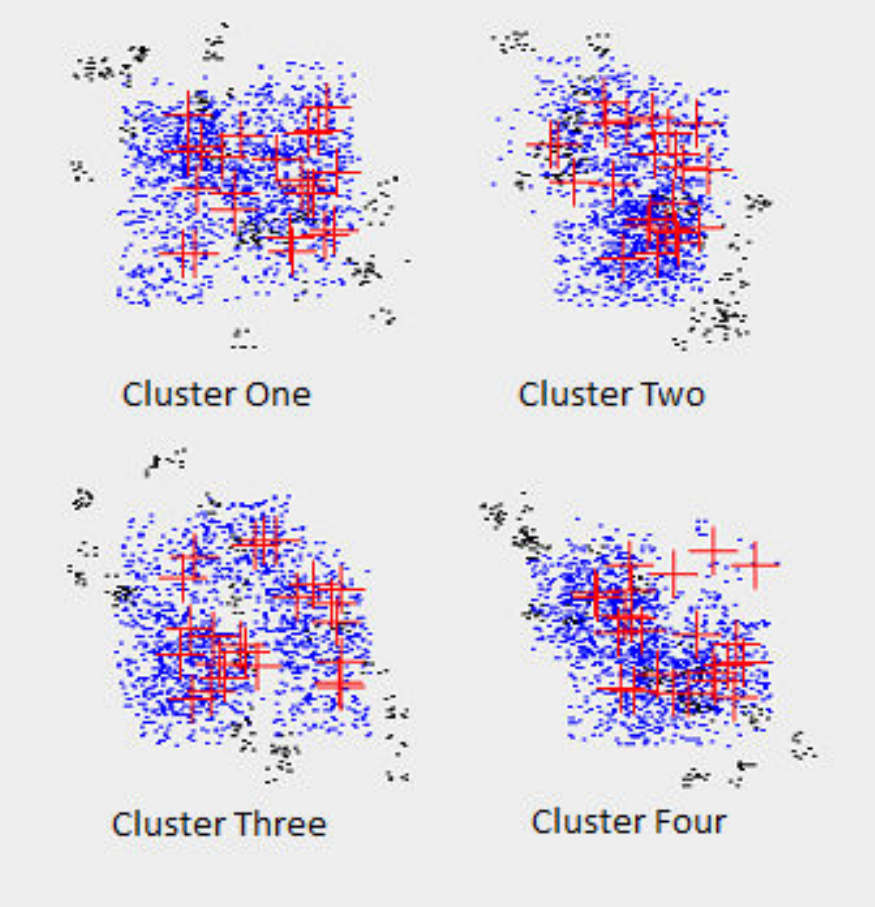}
\caption{Visualizing samples (abstracts) of four clusters of RFC dataset. Red-cross shows abstract terms of a cluster, blue and black dots, respectively, show terms whose topics are covered and not covered by the abstract terms.}
\label{fig:samplingQuality}
\end{figure}
\subsubsection{Quality of Clusters' Sampling in Abstracts}
The quality of sampled terms (\ie abstracts) depends on how well those terms cover the topics within the whole cluster population. To evaluate the quality of sampling from the clustered terms, in Figure~\ref{fig:samplingQuality}, we visually show the dispersion of sampled terms (red-cross symbols in the figure) throughout a cluster. In addition, we show the terms whose topics are captured with the abstract terms (shown in blue dots) and those topics that are not captured (shown in black dots). 

As we can see in this figure, most of the terms in the cluster are represented by the abstract terms and the abstract terms are well distributed throughout the clusters. 

\subsubsection{Evaluating Pruning Accuracy}
\begin{figure}
\centering
\includegraphics[width=0.35\textwidth]{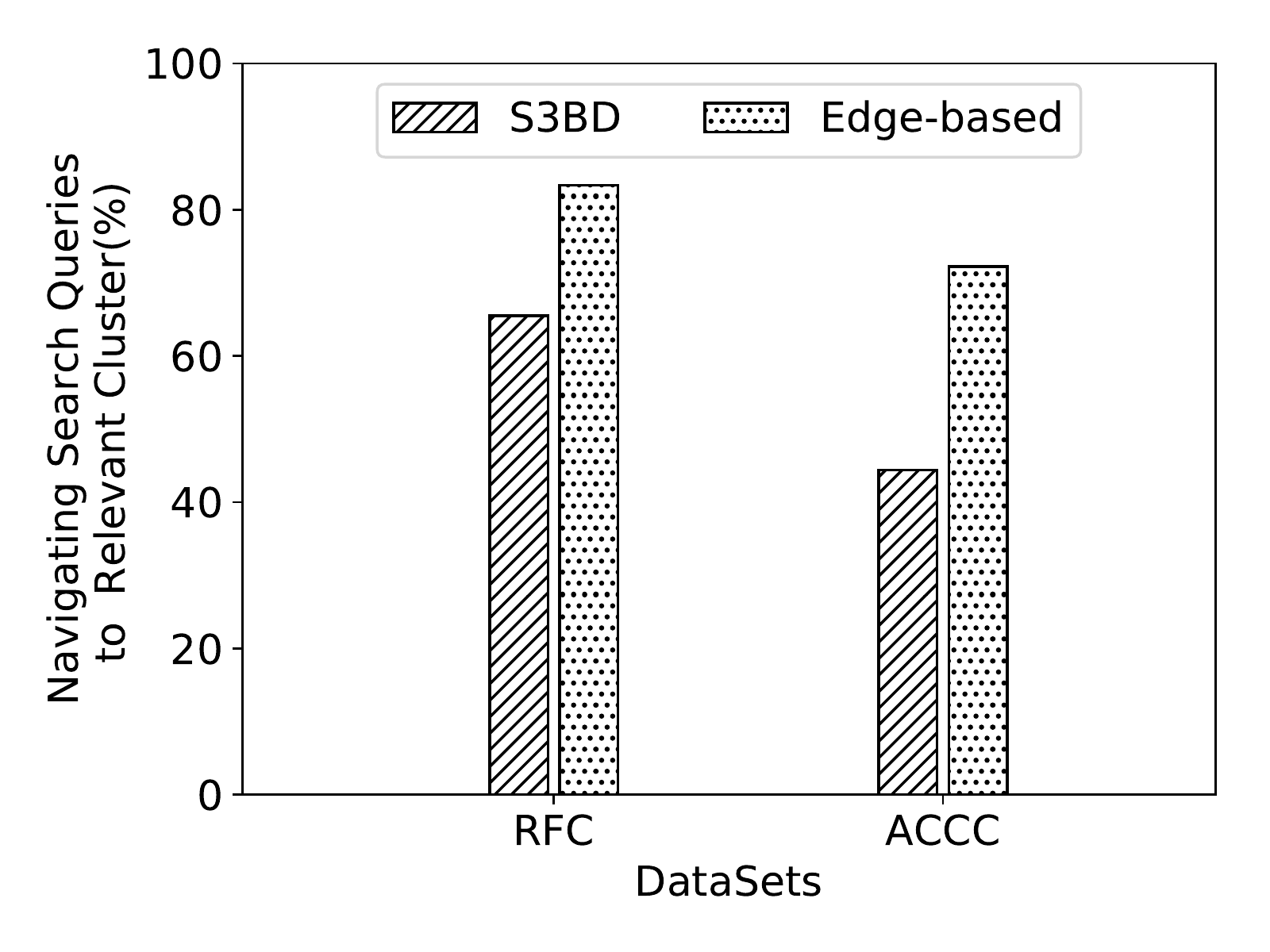}
\caption{Pruning accuracy of edge-based approach versus S3BD. The vertical axis shows the percentage of benchmark queries navigated to relevant clusters and the horizontal axis represents the evaluated dataset.}
\label{fig:SR2}
\end{figure}

The sampling (\ie abstract) quality from clusters is crucial for accurate pruning and targeting relevant clusters to search. To evaluate pruning accuracy, we run the benchmark queries and measure the percentage queries that hit the relevant clusters and compare it with the original approach in S3BD. The result of this experiment is shown in Figure~\ref{fig:SR2} for different datasets. We observe that, for both RFC and ACCC datasets, our approach remarkably outperforms the original S3BD approach in pruning and navigating the search to relevant clusters. The difference between our edge-based approach and S3BD is more remarkable for ACCC dataset, because this dataset is not domain-specific and its topics are more diverse, so that they cannot be captured by only sampling from highly associated documents. %However, in the batch search queries, there are queries, those have less number of document association in the ACCC dataset, as a result, the S3BD abstract can cover an only limited number of topics.

We also observe that pruning accuracy of the edge-based approach for the ACCC dataset is lower than the RFC dataset. The reason is that, in ACCC, there are numerous noun phrases (\eg for location and institution names) that has no general meaning. However, to keep generality in the abstracts, we opted out those specific terms. Hence, when such terms are searched, the pruner cannot navigate the search to relevant clusters. We can conclude that our edge-based approach particularly performs well in sampling general terms as opposed to specific ones.

\subsubsection{The Impact of Abstract Size on Pruning Accuracy}
The proposed abstract manager determines both the quality and quantity of terms to be sampled in abstracts. In this experiment, we concentrate on the quantity aspect of sampling, controlled by the semantic radius. We note that larger samples generally increase the pruning accuracy, however, they increase the overhead too. As such, in this experiment, we study the impact of different ways to determine semantic radius and their respective impacts on both the pruning accuracy and overhead.
We name the semantic radius mentioned in equation~\ref{eq:SR} as the \emph{edge-based} approach and compare it against two other approaches for the semantic radius: (A) user interest ($\beta$); and (B) cluster size plus variation of topics ($\gamma+\delta$). 

\begin{figure}
\centering
\includegraphics[width=0.35\textwidth]{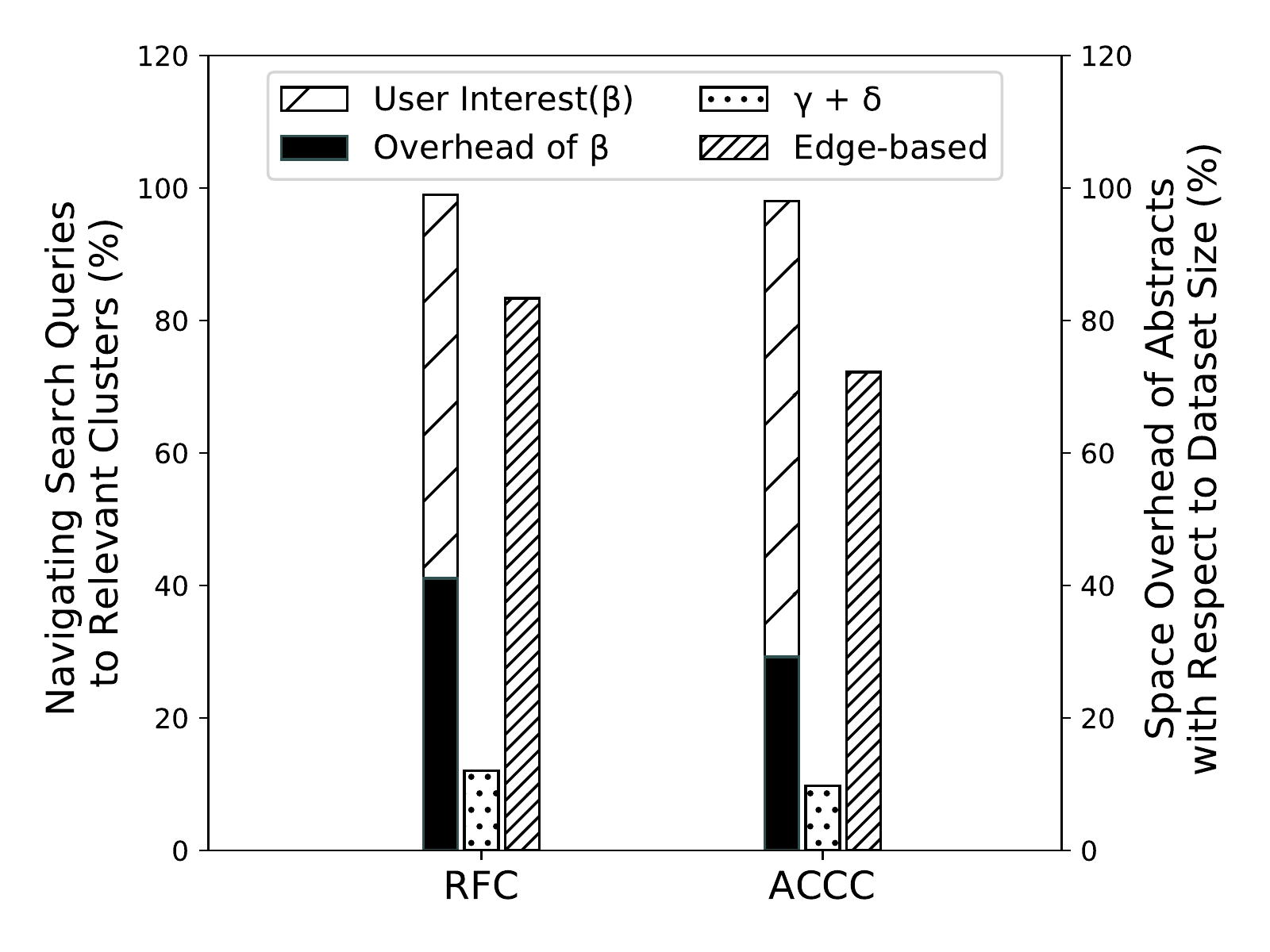}
\caption{The impact of abstract size in RFC and ACCC datasets on pruning accuracy and abstracts' overhead. The vertical axis shows both the percentage of queries navigated to relevant clusters and the abstracts' overhead.}
\label{fig:ABtesting}
\end{figure}

As shown in Figure~\ref{fig:ABtesting}, for different datasets, we measure the percentage of benchmark queries navigated to relevant clusters. We also measure the abstracts' overhead based on the sample size to the total number of dataset terms.  

% We have tested the boundary condition in RFC and ACCC datasets. The boundary case is, \textit{Smaller clusters receive higher user search request (7 small clusters, 3 larger clusters)}.

In RFC and ACCC datasets, using user interest ($\beta$) semantic radius leads to $\approx$99\% pruning accuracy, however, it creates prohibitively large abstracts ($\approx$41.07\% and $\approx$29.23\% of the dataset terms) that cannot be called samples anymore. %In fact, considering only the user cluster into the cluster, so the $\beta$ gain higher value. 
%As a result, the $SR$ of $\beta$ goes to minimum (close to zero). 
%So, almost every search terms lead to the small clusters are being added to the abstracts. 
In case of $\gamma + \delta$, we observe that the pruning accuracy is unacceptably low, because the abstract sizes are small and are not representative of the clusters' topics. Alternatively, in the edge-based approach, we observe that the pruning accuracy is relatively high and the overhead is negligible ($\approx$0.02\%).

\subsubsection{Analyzing Search Time and Space Overhead of the Edge-Based Approach}

%Search precision helps to understand the relevance of document return by the system. In order to measure the search efficiency, we manually check the search precision against the random query. We estimate the search precision up to top 10 ranked document. The search precision of RFC dataset is .... and ACCC dataset is ...... based on result write something
% \begin{figure}
% \centering
% \includegraphics[width=0.3\textwidth]{images/SearchTimeProcessTime.pdf}
% \caption{}
% \label{fig:searchTimeProcess}
% \end{figure}

\begin{figure}
\centering
\includegraphics[width=0.35\textwidth]{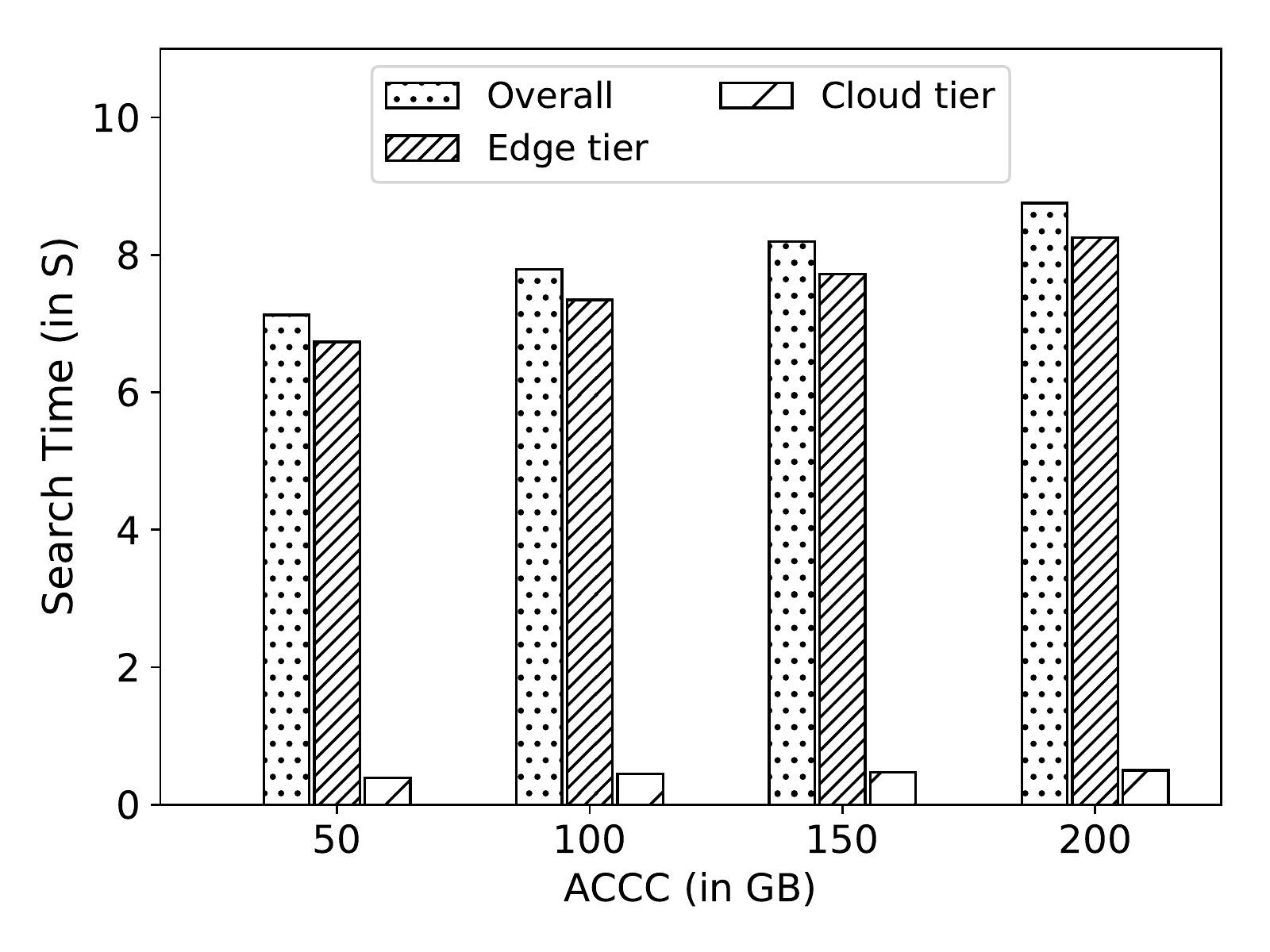}
\caption{Analyzing search time overhead of edge-based system. Overall search time, the time overhead for edge tier and cloud tier are reported. Vertical axis shows time (in S) and horizontal axis shows  subsets of the ACCC dataset with different sizes (in GB). }
\label{fig:SearchTimeAnalysis}
\end{figure}

Although we showed that the proposed edge-based approach improves pruning accuracy and, subsequently, the search precision, the approach comes with an overhead. In this experiment, we elaborate on details of the induced overheads, both in terms of the search time and occupied space. 
The overall Search time in the edge-based approach is dominated by the time to search abstracts on the edge tier and the time to search clusters on the cloud tier. 
As such, to evaluate the search time overhead in this experiment, for each benchmark query, we measure the overall search time, and the overhead time on the edge and cloud tiers. To realize the overhead growth, we examined datasets with various volumes. We used ACCC dataset and created subset datasets ranging from 50 GB to 200 GB. To cluster these datasets, we used the method implemented in S3BD for clustering encrypted data. We created 100 benchmark queries that include one to three keywords and run them against the created datasets. Then, we measured the average of the overall search times, plus the average overhead time of on the edge and cloud tiers. The result of this experiment is shown in Figure~\ref{fig:SearchTimeAnalysis}. In this figure, the horizontal axis shows the size of different datasets created in GB. The vertical axis shows both the overall search time and the overhead of edge tier (in seconds).

Figure~\ref{fig:SearchTimeAnalysis} shows the average of overall search time taken to complete the search queries. We observe that as the dataset size increases, the overall search time also increases. We also observe that the overall search time is dominated by the edge tier and not the cloud tier. The reason is that in S3BD, always three clusters are chosen to be searched, however, to find those three clusters, all abstracts must be checked against each search query. As the size of datasets increases, the number of terms in abstracts also increases, hence, the pruning operation dominates the search time.
%The query augmentation time, the pruning time on the edge tier increase at $\approx$1.0106$X$ rate as more terms added to the abstract to cover more sub-topics for the additional data volume. However, after pruning decision, the system searches the specific relevant cluster, not the entire dataset and reduces the overall search completion time for the additional volume of data as the cloud search completion time increases at $\approx$.0572$X$ rate.

\begin{figure}
\centering
\includegraphics[width=0.35\textwidth]{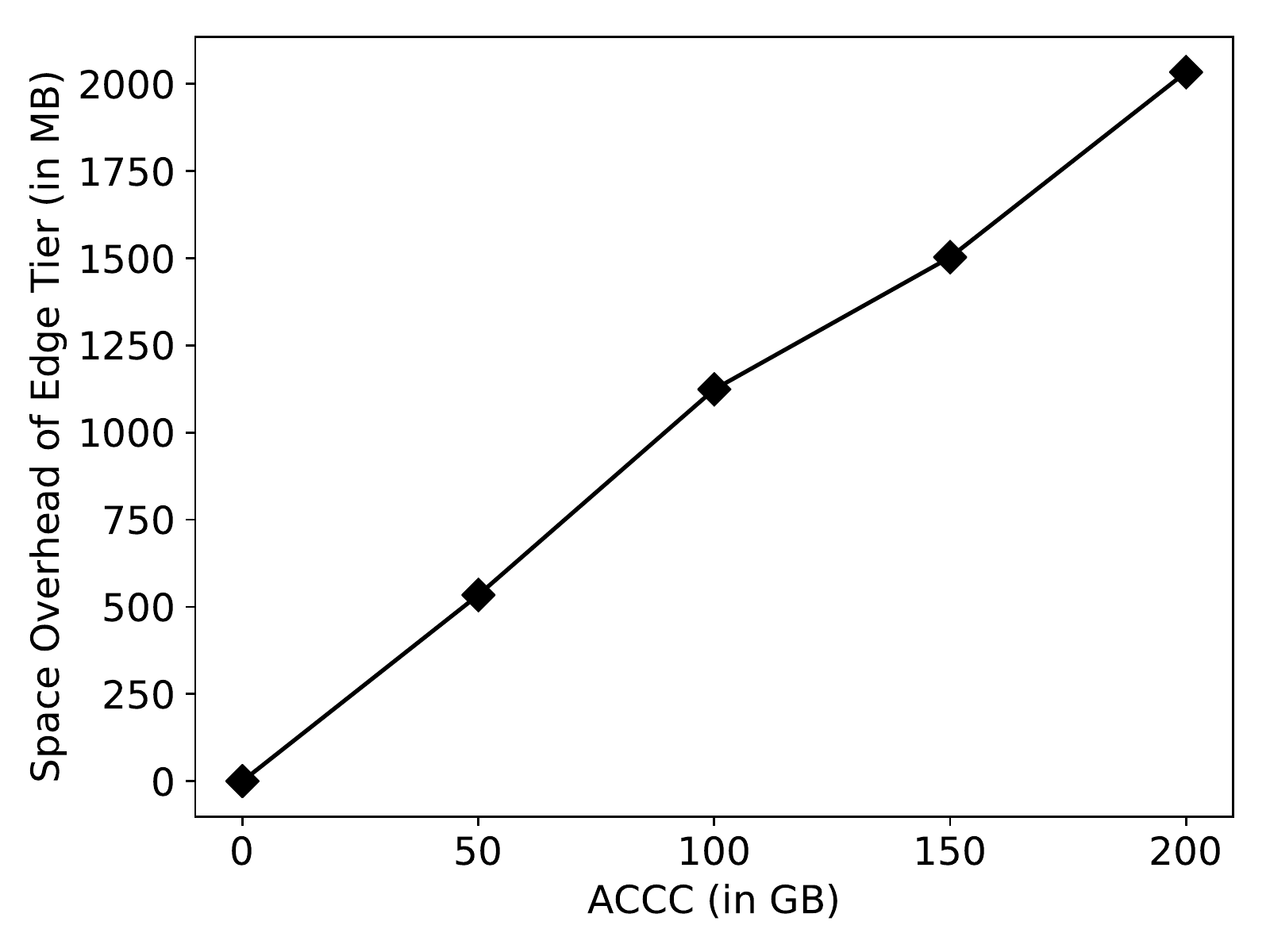}
\caption{Analyzing memory overhead of edge tier for datasets with different volumes. Vertical axis shows memory overhead of the edge tier (in MB) and horizontal axis shows subsets of the ACCC dataset with different sizes (in GB).}
\label{fig:edgeOverhead}
\end{figure}

In the second part of this experiment, we analyze the space overhead of the edge-based approach. The only space overhead imposed in this approach is in the edge tier, thus, we measured memory consumption in the edge tier for various dataset sizes (ranging from 50 GB to 200 GB). The result of this analysis is presented in Figure~\ref{fig:edgeOverhead}.

We observe that there is a linear relationship between the dataset size and the memory overhead of the edge tier  because of the space consumed to maintain the users' search history that is used to analyze their search patterns and constructing effective abstracts. We note that, the space overhead of the edge tier is only about $\approx$0.01\% of the size of the dataset stored in the cloud.

\section{Conclusion and Future Work}\label{sec:conclusion}
In this research, we developed an edge-computing-based framework that offers a user-centric search ability on encrypted big data in cloud. The framework is composed of three tiers namely, user tier, edge tier, and cloud tier. It enables real-time search over encrypted big data by pruning the search space and limiting the search to only relevant clusters of data. For accurate pruning, the edge tier leverages the users' search patterns and creates dynamic samples (aka abstracts) for each cluster. The framework determines quantity of items (terms) in each abstract and populates the abstract with terms that qualitatively represent topics of its corresponding cluster. %We analyze user search tendency on edge tier and based on that analysis along with cloud data volume information, perform sampling operation to create the qualitative and quantitative sample for improving pruning quality which has an impact on search precision.  
We evaluated the pruning quality and navigating the search queries to relevant clusters by comparing the proposed framework against the one used in S3BD search system. Experimental results from different datasets show that the pruning quality is improved by $\approx$27\%. This gain in performance is attained without imposing a major overhead to the system.

There are several avenues that this research can be extended. One avenue will be incorporating the impact of aging on historic users' search queries. Another avenue will be providing a method to evaluate quality of samples created from an encrypted cluster.

\section*{Acknowledgments}
\small{
We thank reviewers of the manuscript. 
This research was supported by the Louisiana Board of Regents under grant number LEQSF(2017-20)-RD-B-06, and Perceptive Intelligence, LLC.} %Lastly, we would like to thank Louisiana Optical Network Infrastructure (LONI) for running our programs in their cluster computer.

%\ifCLASSOPTIONcaptionsoff
%  \newpage
%\fi
%
%\footnotesize
%\linespread{0.01}
\linespread{.872}
 \bibliographystyle{IEEEtran} %cmd 5/21/12 - fix reference format
\bibliography{references}
%\bibcite{amini14}{1}

%\pagebreak
%\input{\paperfolder/Sources/sec-appendix}

\end{document}